\documentclass[preprintnumbers,superscriptaddress,showpacs,pra]{revtex4}%
\usepackage[colorlinks=true,linkcolor=blue]{hyperref}
\usepackage{amssymb}
\usepackage{amsfonts}
\usepackage{amsmath}
\usepackage{graphicx}%
\providecommand{\U}[1]{\protect\rule{.1in}{.1in}}

\let\stdsection\section
\renewcommand\section{\nopagebreak\stdsection}
\begin{document}
\title{General covariant geometric momentum, gauge potential and a Dirac fermion on a
two-dimensional sphere}
\author{Q. H. Liu}
\email{quanhuiliu@gmail.com}
\affiliation{School for Theoretical Physics, College of Physics and Electronics, Hunan
University, Changsha 410082, China}
\affiliation{Synergetic Innovation Center for Quantum Effects and Applications (SICQEA),
Hunan Normal University, Changsha 410081, China}
\author{Z. Li}
\affiliation{School for Theoretical Physics, College of Physics and Electronics, Hunan
University, Changsha 410082, China}
\author{X. Y. Zhou}
\affiliation{School for Theoretical Physics, College of Physics and Electronics, Hunan
University, Changsha 410082, China}
\author{Z. Q. Yang}
\affiliation{School for Theoretical Physics, College of Physics and Electronics, Hunan
University, Changsha 410082, China}
\author{W. K. Du}
\affiliation{School for Theoretical Physics, College of Physics and Electronics, Hunan
University, Changsha 410082, China}
\date{\today}

\begin{abstract}
For a particle that is constrained on an ($N-1$)-dimensional ($N\geq2$) curved
surface, the Cartesian components of its momentum in $N$-dimensional flat
space is believed to offer a proper form of momentum for the particle on the
surface, which is called the geometric momentum as it depends on the mean
curvature. Once the momentum is made general covariance, the spin connection
part can be interpreted as a gauge potential. The present study consists in
two parts, the first is a discussion of the general framework for the general
covariant geometric momentum. The second is devoted to a study of a Dirac
fermion on a two-dimensional sphere and we show that there is the
generalized\textit{ }total angular momentum whose three cartesian components
form the $su(2)$ algebra, obtained before by consideration of dynamics of the
particle, and we demonstrate that there is no curvature-induced geometric
potential for the fermion.

\end{abstract}

\maketitle

\section{Introduction}

In quantum mechanics, there are fundamental quantum conditions (FQCs)
$[x_{i},x_{j}]=0,$ $[x_{i},p_{j}]=i\hbar\delta_{ij},$ and $[p_{i},p_{j}]=0$,
which are defined by the commutation relations between positions $x_{i}$ and
momenta $p_{i}$ ($i,j,k,l=1,2,3,...,N$) where $N$ denotes the number of
dimensions of the flat space in which the particle moves \ \cite{Dirac}. In
position representation, the momentum operator takes simple form as
$\mathbf{p}=-i\hslash\nabla$ where $\nabla\equiv\mathbf{e}_{i}\partial
/\partial x_{i}$ is the ordinary gradient operator, and $N$ mutually
orthogonal unit vectors $\mathbf{e}_{i}$ span the $N$ dimensional Euclidean
space $E^{N}$. Hereafter the Einstein summation convention over repeated
indices is used. When the particle is constrained to remain on a hypersurface
$\Sigma^{N-1}$ embedded in $E^{N}$, the FQCs become\ \ \cite{weinberg},%
\begin{equation}
\lbrack x_{i},x_{j}]=0,\text{ }[x_{i},p_{j}]=i\hbar(\delta_{ij}-n_{i}%
n_{j}),\text{ and }[p_{i},p_{j}]=-i\hbar\left\{  ({n_{i}}{n_{k}}_{,j}-{n_{j}%
}{n_{k}}_{,i})p_{k}\right\}  _{Hermitian},\label{fqc}%
\end{equation}
where $O_{_{Hermitian}}$ stands for a Hermitian operator of an observable $O$,
and the equation of surface $f(\mathbf{x})=0$ can be so chosen that
$\left\vert \nabla f(\mathbf{x})\right\vert =1$ so $\mathbf{n\equiv}\nabla
f(\mathbf{x})=$ $\mathbf{e}_{i}n_{i}$ being the normal at a local point on the
surface. This set of the FQCs\textit{ }(\ref{fqc}) is highly non-trivial, from
which it is in general impossible to uniquely construct the momenta $p_{i}$.
Our propose of the proper form of the momentum for a spinless particle was
\cite{liu07,liu11,liu13,liu132,liu18},
\begin{equation}
{\mathbf{p}}=-i\hbar({\nabla_{\Sigma}}+{M{\mathbf{n/}}}2),\label{ver1}%
\end{equation}
where ${\nabla_{\Sigma}\equiv}\mathbf{e}_{i}(\delta_{ij}-n_{i}n_{j}%
)\partial_{j}$ $=\nabla-\mathbf{n}\partial_{n}$ $=\mathbf{r}^{\mu}%
\partial_{\mu}$ is the the gradient operator, and $\mathbf{r}^{\mu}$ is $\mu
-$th contravariant component of the natural frame on the point ($u^{1}%
,u^{1},...u^{\mu},...u^{N-1}$) on the surface $\Sigma^{N-1}$, and $u^{\mu}$
($\mu,\nu,\alpha,\beta=1,2,3,...,N-1$) denote the local coordinates, and the
mean curvature ${M\equiv-\nabla_{\Sigma}}\cdot\mathbf{n}$ is defined by the
sum of the all principal curvatures. Since the mean curvature ${M}$ is an
extrinsic curvature, this form of momentum (\ref{ver1}) is fundamentally
different from the canonical ones in curvilinear coordinates for it depends on
the geometric invariants. Thus it can be conveniently called as geometric
momentum \cite{liu07,liu11,liu13,liu132,liu18,exp,wang17}. This momentum can
be obtained by many different ways including: the hermiticity requirement on
derivative part $-i\hbar{\nabla_{\Sigma}}$ \cite{liu07}, and compatibility of
constraint condition $\mathbf{n\cdot p+p\cdot n}=0$ which means that the
motion is perpendicular to the surface normal vector $\mathbf{n}$
\cite{liu11,liu13}, and thin-layer quantization or confining potential
formalism which instead considers that particle is confined onto the surface
$\Sigma^{N-1}$ by means of introduction of a confinement potential along the
normal direction of the the surface \cite{liu132}, and dynamical quantum
conditions (DQCs) \cite{liu18}, etc. \cite{wang17,ikegami} It was demonstrated
that this momentum (\ref{ver1}) satisfies last one of the FQCs (\ref{fqc}),
when it explicitly takes following simplest form \cite{liu18},%
\begin{equation}
\lbrack p_{i},p_{j}]=-i\hbar\left\{  ({n_{i}}{n_{k}}_{,j}-{n_{j}}{n_{k}}%
_{,i})p_{k}+p_{k}({n_{i}}{n_{k}}_{,j}-{n_{j}}{n_{k}}_{,i})\right\}
/2\label{sym}%
\end{equation}
Experimental justification was performed by comparison of the interference
spots formed by the surface plasmon polariton propagating on a cylindrical
surface, predicted by the introduction of the geometric momentum or not
\cite{exp}, respectively. Some of previous discussions deal with quite general
case \cite{liu13,liu18,ikegami}, some of them \cite{liu07,liu11,liu132,wang17}
are mainly for a particle on $\Sigma^{2}$.

The geometric momentum (\ref{ver1}) suffices to act on state function that has
a single component. However, state functions on the surfaces are usually
multi-component such as spinors \cite{RMP1957,2010PR,LeeDH,Iorio}, requiring
that the momentum be made general covariance. In fact, the general covariant
geometric momentum (GCGM) is at hand, though not yet explicitly written
before. The present paper shows that the GCGM is a useful and convenient
physical quantity.

This paper is mainly divided into two parts. Sections II-IV are devoted to
build up a general formulation between the GCGM and quantization conditions.
Section V and VI study the Dirac fermion on $S^{2}$. In section II, the
introduction of the GCGM is made and its dependence on the gauge potential is
transparent. In section III, though we do not know in general whether the GCGM
satisfies the quantization condition (\ref{sym}), or satisfies other forms of
the last one of the FQCs (\ref{fqc}) $[p_{i},p_{j}]=-i\hbar\left\{  ({n_{i}%
}{n_{k}}_{,j}-{n_{j}}{n_{k}}_{,i})p_{k}\right\}  _{Hermitian}$, the FQCs
$\left[  p_{i},p_{j}\right]  $ for a Dirac fermion on $S^{N-1}$ have a
well-defined consequence to define a \emph{generalized total angular
momentum}. In section IV, we show how the self-consistent consideration of the
quantization conditions leads us to the DQCs for a relativistic particle on
$\Sigma^{N-1}$. In section V, we deal with the Dirac fermion on $S^{2}$, and
use FQCs and GCGM to reproduce the same \emph{generalized total angular
momentum} obtained before by means of a purely dynamical consideration. In
section VI, we use DQCs and GCGM to check whether the curvature-induced
geometric potential presents for a Dirac fermion on $S^{2}$, and results show
that no such a potential. Final section VII is a brief conclusion.

\section{General covariant geometric momentum and gauge potential}

This section is to show that the GCGM is at the ready, and its dependence on
the gauge potential is transparent. 

To note ${\nabla_{\Sigma}\equiv}\mathbf{e}_{i}(\delta_{ij}-n_{i}n_{j}%
)\partial_{j}=\mathbf{r}^{\mu}\partial_{\mu}$, and the usual derivative
$\partial_{\mu}(\equiv\partial/\partial u^{\mu})$ in (\ref{ver1}) can be made
general covariant by a simple replacement \cite{RMP1957,2010PR,LeeDH,Iorio},%
\begin{equation}
\partial_{\mu}\longrightarrow D_{\mu}\equiv\partial_{\mu}+i\Omega_{\mu
}\label{rep1}%
\end{equation}
and we immediately have,
\begin{equation}
{\mathbf{p}}=-i\hbar({\nabla_{\Sigma}}+\frac{{M{\mathbf{n}}}}{2}%
+i\mathbf{r}^{\mu}\Omega_{\mu})=-i\hbar({\nabla_{\Sigma}}+\frac{{M{\mathbf{n}%
}}}{2})+\hbar\mathbf{r}^{\mu}\Omega_{\mu},\label{GM}%
\end{equation}
where $\Omega_{\mu}=\left(  -i/8\right)  \omega_{\mu}^{ab}\left[  \gamma
_{a},\gamma_{b}\right]  $ in which $\omega_{\mu}^{ab}$ are the spin
connections \cite{2010PR,LeeDH,Iorio,ogawa} and $\gamma_{a}$ ($a,b=0,1,2,...N$%
) are Dirac spin matrices. In comparison of (\ref{GM}) with the usual
kinematical momentum ${\mathbf{p}}=-i\hbar{\nabla-q}\mathbf{A}$ in presence of
magnetic potential $\mathbf{A}$, we see that an equivalent magnetic potential
$\mathbf{A}$ can be defined by ${q}\mathbf{A\equiv}{-}\hbar\mathbf{r}^{\mu
}\Omega_{\mu}$, in which the charge ${q}$ can be understood as an effective
interaction strength between the charge with the field. Once writing
$\Omega_{\mu}$ as a product of $\omega_{\mu}^{ab}/4$ and $Q_{ab}\equiv\left[
\gamma_{a},\gamma_{b}\right]  /(2i)$, we can take the eigenvalues of the
matrices $Q_{ab}$ as an effective interaction strength \cite{ogawa}. This form
of GCGM (\ref{GM}) is applicable to particles, relativistically or not,
massively or not.

Two observations concerning the GCGM are in following.

1. Once the surface $\Sigma^{N-1}$ is embedded into higher flat space in
$E^{N+p}$ ($N\succeq2$) with a positive integer $p\succeq1$, we have another
way of making the derivative $\partial_{\mu}$ in (\ref{ver1}) covariant by
replacement \cite{ogawa,Nconnection,maraner},%
\begin{equation}
\partial_{\mu}\longrightarrow D_{\mu}\equiv\partial_{\mu}+iW_{\mu}\label{rep2}%
\end{equation}
and we have as well,
\begin{equation}
{\mathbf{p}}=-i\hbar({\nabla_{\Sigma}}+\frac{{M{\mathbf{n}}}}{2}%
+i\mathbf{r}^{\mu}W_{\mu})=-i\hbar({\nabla_{\Sigma}}+\frac{{M{\mathbf{n}}}}%
{2})+\hbar\mathbf{r}^{\mu}W_{\mu},\label{ver3}%
\end{equation}
where $\hbar W_{\mu}=N_{\mu}^{AB}L_{AB}/2$ in which $N_{\mu}^{AB}$
($A,B=N,N+1,...N+p$) stand for the normal connections determined by the
so-called normal fundamental form, and $L_{AB}$ are angular momentum in the
normal space. It was realized that the normal connections $N_{\mu}^{AB}/2$ and
spin connections $\omega_{\mu}^{ab}/4$ can take identical form for $S^{N-1}$,
which was used to explore an origin of spin other than that is generally
accepted to be connected with relativity \cite{OK}. Thus, the relationship
between spin and space embedding is far from fully understood.

2. Starting from replacement (\ref{rep1}) or (\ref{rep2}), we can define the
gauge potential $A_{\mu}\equiv-\hbar\Omega_{\mu}$, or $A_{\mu}\equiv-\hbar
W_{\mu}$. Therefore the field strength $F_{\mu\nu}$ can be defined by
\ \cite{Nconnection,maraner,ogawa,ohnuki},%
\begin{equation}
F_{\mu\nu}\equiv\partial_{\mu}A_{\nu}-\partial_{\nu}A_{\mu}-i[A_{\mu},A_{\nu
}]=i[D_{\mu},D_{\nu}].\label{FS}%
\end{equation}
Whether this gauge field $F_{\mu\nu}$ is abelian or non-abelian depends on
whether the commutators $[A_{\mu},A_{\nu}]$\ vanish or not. In terms of the
GCGM, we have the gauge potential in Cartesian coordinates,
\begin{equation}
\mathbf{A}\equiv\mathbf{r}^{\mu}A_{\mu}=-\hbar\mathbf{r}^{\mu}\Omega_{\mu
},\text{ or }-\hbar\mathbf{r}^{\mu}W_{\mu}.\label{gaugefield}%
\end{equation}

Now the introduction of the GCGM is complete. However whether it satisfies the
quantization condition (\ref{sym}), or satisfies other forms of the last one
of the FQCs (\ref{fqc}) $[p_{i},p_{j}]=-i\hbar\left\{  ({n_{i}}{n_{k}}%
_{,j}-{n_{j}}{n_{k}}_{,i})p_{k}\right\}  _{Hermitian}$, is not so easily
resolved in general. We leave it as an open problem though we believe it is
true. For the special case of a Dirac fermion on $S^{N-1}$, this problem turns
out to be another one defining instead the \emph{generalized total angular
momentum}, which will be discussed in next section.

\section{Fundamental quantum conditions $\left[  p_{i},p_{j}\right]  $ for a
Dirac fermion on $S^{N-1}$}

The hypersphere $S^{N-1}$ of radius $R$ in $N$-dimensional flat space $E^{N}$
can be,%
\begin{equation}
f(\mathbf{x})\equiv\frac{1}{2R}\left(  \sum_{i=1}^{N}x_{i}^{2}-R^{2}\right)
=0.\label{hypersphere}%
\end{equation}
The fundamental set of Dirac brackets is simply \cite{weinberg,liu11,liu13},%
\begin{equation}
\left[  x_{i},x_{j}\right]  _{D}=0,\text{ }\left[  x_{i},p_{j}\right]
_{D}=\left(  \delta_{ij}-n_{i}n_{j}\right)  ,\text{ and }\left[  p_{i}%
,p_{j}\right]  _{D}=-\frac{L_{ij}}{R^{2}},\label{fdb}%
\end{equation}
where $L_{ij}\equiv x_{i}p_{j}-x_{j}p_{i}$ is the $ij$-component of the
\textit{orbital} angular momentum. In addition, we have an $SO(N,1)$ group
with generators $p_{i}$ and ${L_{ij}}$ because we have also
\ \cite{liu13,ohnuki},
\begin{equation}
\lbrack{L_{ij}},{L_{k\ell}}]_{D}={-{\delta_{i\ell}}{L_{kj}+{\delta_{ik}%
}{L_{\ell j}}+\delta_{jk}}{L_{i\ell}}-{\delta_{j\ell}}{L_{ik}}},\text{ and
}\left[  {{L_{ij}},p{_{\ell}}}\right]  _{D}=\left(  {{\delta_{i\ell}}p{_{j}%
}-{\delta_{j\ell}}p{_{i}}}\right)  .\label{son1}%
\end{equation}
These relations (\ref{fdb}) and (\ref{son1}) hold irrespective of particle
being massive or not, relativistic or not. However, in classical mechanics,
there is no spin; and these relations (\ref{son1}) are obtained by considering
the purely orbital motion. In our approach, we require that these relations
(\ref{fdb}) and (\ref{son1}) hold true in sense of $\left[  u,v\right]
=i\hbar O\left(  \left[  u,v\right]  _{D}\right)  $. Explicitly, we have,%
\begin{equation}
\left[  p_{i},p_{j}\right]  =-i\hbar\frac{J_{ij}}{R^{2}},\text{ }\left[
J{{_{ij}},p{_{\ell}}}\right]  =i\hbar\left(  {{\delta_{i\ell}}p{_{j}}%
-{\delta_{j\ell}}p{_{i}}}\right)  ,\text{ and }[J{_{ij}},J{_{k\ell}}]=i\hbar
O\left(  [J{_{ij}},J{_{k\ell}}]_{D}\right)  .\label{SON1}%
\end{equation}
Here we re-denote $L_{ij}$ by the symbol $J{_{ij}}$, a symbol denoting
\emph{generalized total angular momentum} in quantum mechanics. Our discussion
needs a flat space with $N$ cartesian coordinates $x_{i}$ ($i=1,2,3,...,N$) as
the prerequisite. So, for a Dirac fermion on $S^{N-1}$, the FQCs are set up
and given by (\ref{SON1}), which lead us to defining the \emph{generalized
total angular momentum} $J$ in quantum mechanics.

When quantizing a classical system, we put symmetries on the top priority:
\cite{liu11,liu13,liu18,liu133,liu15} Our philosophy is: \emph{The symmetry
expressed by the Poisson or Dirac brackets in classical mechanics preserves in
quantum mechanics; and so the Hamiltonian is determined by the symmetry. }It
can be considered a specific demonstration of the fundamental philosophical
idea stating that \emph{symmetry dictates interactions} \emph{in quantum
mechanics} \cite{yang}. The philosophy leads us to set out FQCs and DQCs for
the non-relativistic and spinless particle, and the most profound consequence
is to successfully reproduce of the geometric potential in Hamiltonian and the
geometric momentum \cite{liu18}, respectively. In next section, the DQCs
affecting the form of Hamiltonian for a relativistic particle on $\Sigma
^{N-1}$ will be formulated.

\section{Dynamical quantum conditions for a relativistic particle on
$\Sigma^{N-1}$}

For a relativistic particle whose classical Hamiltonian is $H=\sqrt{\left(
pc\right)  ^{2}+\left(  \mu c^{2}\right)  ^{2}}$ with $c$ being the velocity
of light and $\mu$ being the mass of the particle, we can obtain two Dirac
brackets,
\begin{equation}
\left[  x_{i},H\right]  _{D}=\frac{p_{i}}{H}c^{2},\text{ and\ }\left[
p_{i},H\right]  _{D}=-n_{i}\kappa\frac{\left(  cp\right)  ^{2}}{H}%
,\label{xphd}%
\end{equation}
where $\kappa$ is the\ first curvature of the geodesic on the hypersurface
$\Sigma^{N-1}$ \ \cite{liu16}. Notice that Eqs. (\ref{xphd}) have two
important consequences,
\begin{equation}
p_{i}=\frac{1}{c^{2}}H\left[  x_{i},H\right]  _{D},\text{ and\ }%
\mathbf{n}\wedge\left[  \mathbf{p},H\right]  _{D}=0.\label{xphd2}%
\end{equation}
These two relations indicate that in quantum mechanics momentum $\mathbf{p}$
and Hamiltonian $H$ must be compatible with following two quantum conditions,%
\begin{equation}
p_{i}=\frac{1}{i\hbar}\frac{H\left[  x_{i},H\right]  +\left[  x_{i},H\right]
H}{2c^{2}},\text{ and\ }\mathbf{n}\wedge\left[  \mathbf{p},H\right]  -\left[
\mathbf{p},H\right]  \wedge\mathbf{n}=0.\label{qph}%
\end{equation}
These two sets of quantum conditions constitute the so-called DQCs for the
relativistic particle on $\Sigma^{N-1}$, which put requirement on the form of
Hamiltonian operator.

Three remarks concerning the DQCs are in following.

1. In classical mechanics for a particle, constrained or not, the relativistic
velocity $\mathbf{v}\equiv\mathbf{p}c^{2}/H$ (\ref{xphd}) can be rewritten as
the familiar form, $p=\mu v/\sqrt{1-v^{2}/c^{2}}$. In quantum mechanics, the
DQCs imply a definition of the velocity operator $\mathbf{v}\equiv
c\mathbf{\alpha}=\left(  H^{-1}\mathbf{p+p}H^{-1}\right)  c^{2}/2$
\ \cite{position}\ with $\mathbf{\alpha}$ are $4\times4$ Pauli matrices and in
Pauli-Dirac representation we have $\mathbf{\alpha=}\left(
\begin{array}
[c]{lc}%
0 & \mathbf{\sigma}\\
\mathbf{\sigma} & 0
\end{array}
\right)  $, while the momentum $\mathbf{p}$ is defined as $\mathbf{p}%
\equiv\left(  H\left[  \mathbf{x},H\right]  +\left[  \mathbf{x},H\right]
H\right)  /(2i\hbar c^{2})$ which is identical to $-i\hslash\nabla$ for motion
in flat space. However, it is not the case in quantum mechanics once the
motion is constrained. In the quantum mechanics, the relativistic Hamiltonian
operator $H$ for a particle of any spin in flat space can be easily
constructed and it acts on the multi-component wave functions. However, the
construction of such a Hamiltonian for a spin particle on a curved space or
curved space is not an easy task at all. Fortunately, such a Hamiltonian for a
Dirac fermion on $S^{2}$ is easily found \cite{2010PR,LeeDH,Iorio,Abrikosov}.
For a Dirac fermion on $S^{2}$, $\mathbf{n}\wedge\left[  \mathbf{p},H\right]
-\left[  \mathbf{p},H\right]  \wedge\mathbf{n}=0$ (\ref{qph}) clearly leads to
no presence of geometric potential, as discussed in section VI.

2. For a spinless particle that moves non-relativistically, two Dirac brackets
(\ref{xphd2}) become $[\mathbf{x},H]_{D}\equiv\mathbf{p/}\mu$ and\ $\mathbf{n}%
\wedge\left[  \mathbf{p},H\right]  _{D}=0$. DQCs take following forms
\cite{liu11,liu13,liu15,liu18},
\begin{equation}
\lbrack\mathbf{x},H]\equiv i\hbar\frac{\mathbf{p}}{\mu};\text{ }%
\mathbf{n}\wedge\lbrack\mathbf{p},H]-[\mathbf{p},H]\wedge\mathbf{n}=0.
\label{qc}%
\end{equation}
Quantum conditions (\ref{fqc}) and (\ref{qc}) constitute the so-called
\textit{enlarged canonical quantization scheme }which gives the unambiguous
forms of both the momentum and Hamiltonian $H$\ for a
\textit{non-relativistic} free particle \cite{liu18,liu15},
\begin{equation}
{\mathbf{p}}=-i\hbar({\nabla_{\Sigma}}+\frac{{M{\mathbf{n}}}}{2});\text{
}H=-\frac{\hbar^{2}}{2\mu}\nabla_{LB}^{2}+V_{G}, \label{pandh}%
\end{equation}
where $\nabla_{LB}^{2}={\nabla_{\Sigma}}\cdot{\nabla_{\Sigma}}$ is the usual
Laplace-Beltrami operator on the surface $\Sigma^{N-1}$, and $V_{G}%
\equiv-\frac{\hbar^{2}}{4\mu}K+\frac{\hbar^{2}}{8\mu}{M}^{2}$ is the
celebrated \textit{geometric potential}
\ \cite{jk,dacosta,fc,packet1,exp1,exp2} in which $K$ is in fact the trace of
square of the extrinsic curvature tensor \cite{ikegami}, and ${\mathbf{p}}$ is
very \textit{geometric momentum}\ without spin connection
\cite{liu11,liu13,exp}. Physical consequences resulting from \textit{geometric
potential }and\textit{ geometric momentum }are experimentally confirmed
\cite{exp,exp1,exp2}, and more experimentally testable results are under
explorations \cite{packet1}.

3. In comparison with the overall successes of the DQCs (\ref{pandh}) for a
non-relativistic and spinless particle on $\Sigma^{N-1}$, we can only say that
for a relativistic and spin particle on $\Sigma^{N-1}$ the GCGM and the
Hamiltonian must be simultaneously compatible with the DQCs (\ref{qph}). Since
the GCGM is already given, we must look for the proper form of the
Hamiltonian. It is well-accepted that Hamiltonian for spinless and
non-relativistic particle contains the geometric potential, so it is taken for
granted that there must be some form of the geometric potential in Hamiltonian
for a spin and relativistic particle. Unfortunately, due to the fact that a
full understanding of spin connection is still lacking, we can not help but
deal with a system case by case. In the rest part of the paper, we mainly deal
with the curvature-induced geometric potential for a Dirac fermion on $S^{2}$.
Before it, we give specific FQCs and DQCs.

\section{Generalized total angular momentum for a Dirac fermion on $S^{2}$}

The surface $S^{2}$ of unit radius can be parameterized by,%
\begin{equation}
x=\sin\theta\cos\varphi;\text{ }y=\sin\theta\sin\varphi;\text{ }z=\cos\theta,
\label{xyz}%
\end{equation}
where $\theta$ is the polar angle from the positive $z$-axis with $0\leq
\theta\leq\pi$, and $\varphi$ is the azimuthal angle in the $xy$-plane from
the $x$-axis with $0\leq\varphi<2\pi$. After some lengthy but straightforward
calculations, we can reach a very simple expression for the general covariant
geometric momentum whose three components are given by,%
\begin{equation}
p_{x}=\Pi_{x}+\sigma_{z}\frac{\hbar}{2}\frac{\cos\theta}{\sin\theta}%
\sin\varphi,\text{ }p_{y}=\Pi_{y}-\sigma_{z}\frac{\hbar}{2}\frac{\cos\theta
}{\sin\theta}\cos\varphi,\text{ and }p_{z}=\Pi_{z}, \label{pz}%
\end{equation}
where $\sigma_{z}=\left(
\begin{array}
[c]{lc}%
1 & 0\\
0 & -1
\end{array}
\right)  $ is the $z$-component Pauli matrix, and $\Pi_{i}$ are geometric
momentum (\ref{pandh}) for the particle on $S^{2}$
\ \cite{liu11,liu13,liu15,liu18},
\begin{align}
\Pi_{x}  &  =-i\hbar(\cos\theta\cos\varphi\frac{\partial}{\partial\theta
}-\frac{\sin\varphi}{\sin\theta}\frac{\partial}{\partial\varphi}-\sin
\theta\cos\varphi);\label{hpx}\\
\Pi_{y}  &  =-i\hbar(\cos\theta\sin\varphi\frac{\partial}{\partial\theta
}+\frac{\cos\varphi}{\sin\theta}\frac{\partial}{\partial\varphi}-\sin
\theta\sin\varphi);\label{hpy}\\
\Pi_{z}  &  =i\hbar(\sin\theta\frac{\partial}{\partial\theta}+\cos\theta).
\label{hpz}%
\end{align}
It has been recognized that spin connection can be interpreted in terms of
gauge potential. In GCGM (\ref{pz}), the gauge potential $\mathbf{A}$ is
evidently,
\begin{equation}
\mathbf{A=}\sigma_{z}\frac{\hbar}{2r}(-\frac{\cos\theta}{\sin\theta}%
\sin\varphi,\frac{\cos\theta}{\sin\theta}\cos\varphi,0)=\sigma_{z}\frac{\hbar
}{2}\frac{1}{\sqrt{x^{2}+y^{2}+z^{2}}}(-\frac{zy}{x^{2}+y^{2}},\frac{zx}%
{x^{2}+y^{2}},0), \label{A}%
\end{equation}
in which the radius $r=\sqrt{x^{2}+y^{2}+z^{2}}$ is recovered. The magnetic
strength $\mathbf{B}$ is,
\begin{equation}
\mathbf{B\equiv\nabla\times A=-}\sigma_{z}\frac{\hbar}{2}\frac{\mathbf{e}_{r}%
}{r^{2}}, \label{B}%
\end{equation}
where $\mathbf{e}_{r}\equiv(x,y,z)/r$. Evidently, the magnetic field is
produced by a monopole of unit charge at the center of the sphere $-\delta
(r)$, and the eigenvalues $\pm\hbar/2$ of the Pauli matrix $\sigma_{z}\hbar/2$
are the effective interaction strength.

The FQCs (\ref{SON1}) for a Dirac fermion on $S^{2}$ are explicitly,
\begin{equation}
\left[  p_{i},p_{j}\right]  =-i\hbar\varepsilon_{ijk}\frac{J_{k}}{R^{2}%
},\text{ }\left[  J{{_{i}},p}_{{{j}}}\right]  =i\hbar\varepsilon_{ijk}{p}%
_{k},\text{ and }[J{_{i}},J{_{j}}]=i\hbar\varepsilon_{ijk}J_{k}.\label{fqcs2}%
\end{equation}
These six operators $p_{i}$ (\ref{pz}) and $j_{i}$ (\ref{jz}) constitute all
generators of an $SO(3,1)$ group. In consequence, we have following
\emph{generalized total angular momentum},
\begin{equation}
J_{x}=L_{x}+\sigma_{z}\frac{\hbar}{2}\frac{\cos\varphi}{\sin\theta},\text{
}J_{y}=L_{y}+\sigma_{z}\frac{\hbar}{2}\frac{\sin\varphi}{\sin\theta},\text{
and }J_{z}=L_{z},\label{jz}%
\end{equation}
where ${L}_{{x}}=i\hbar(\sin\varphi\partial_{\theta}+\cot\theta\cos
\varphi\partial_{\varphi})$, ${L}_{{y}}=-i\hbar(\cos\varphi\partial_{\theta
}-\cot\theta\sin\varphi\partial_{\varphi})$ and ${L}_{{z}}=-i\hbar
\partial_{\varphi}$ are usually $x$, $y$ and $z$ -component of the orbital
angular momentum, respectively. This \emph{generalized total angular momentum}
was first constructed explicitly by Abrikosov in 2002, \cite{Abrikosov} who
observed the Hamiltonian for massless Dirac fermion to be invariant under a
$SU(2)$ group transformation and identified (\ref{jz}) as a consequence. Then,
Abrikosov demonstrated it is really \emph{generalized total angular momentum}
\cite{Abrikosov} for the eigenvalues of $J^{2}\equiv J_{i}J_{i}$ are
$j(j+1)\hbar^{2}$ with $j=1/2,3/2,5/2,...$. In other words, Abrikosov obtained
the \emph{generalized total angular momentum} (\ref{jz})\ on the base of
dynamics. In contrast, we obtain the same result (\ref{jz}) from both the FQCs
(\ref{SON1}) and GCGM (\ref{GM}). Moreover, in this section, our result
(\ref{jz}) applies for particle, massive or massless, relativistic or
non-relativistic, irrespective the form of Hamiltonian. In the history, Ohnuki
and Kitakado \cite{ohnuki}\ in 1993 created the so-called fundamental algebra
for quantum mechanics on $S^{N-1}$ and obtained generators of $SO(N)$ which
when $N=3$ reduces to be $J_{x}=L_{x}+Sy/(r^{2}+rz),$ $J_{y}=L_{y}%
-Sx/(r^{2}+rz),$ and $J_{z}=L_{z}$, in which $S$ is a real number rather than
operator in our situation. However, the monopole is the same. Especially,
Ohnuki and Kitakado also considered the momentum operators on $S^{N-1}$ but
they obtained the geometric one (\ref{ver1}) rather than GCGM (\ref{GM}).

\section{No geometric potential for a Dirac fermion on $S^{2}$}

The so-called geometric potential is the additional term in Hamiltonian
resulting from quantization. Recently, whether such a curvature-induced
geometric potential presents is a topic of considerable controversy
\cite{maraner,packet9,jose}, and all use the confining potential formalism but
have opposite results. Our approach based on the DQCs (\ref{qph}) is totally
different from the confining potential formalism, which are transparent and convincing.

The general covariant Dirac equation for a fermion on a two-dimensional sphere
is \cite{2010PR,LeeDH,Iorio},
\begin{equation}
-i\hbar\gamma^{\mu}\left(  \partial_{\mu}+\Omega_{\mu}\right)  \Psi-\gamma
^{0}m\Psi=0.
\end{equation}
where $m\equiv\mu c$ is the reduced mass. The Hamiltonian can be shown to be
given by \cite{2010PR,Abrikosov},
\begin{equation}
H=-i\hbar\left(  \sigma_{x}\left(  \partial_{\theta}+\frac{1}{2}\frac
{\cos\theta}{\sin\theta}\right)  +\sigma_{y}\frac{1}{\sin\theta}%
\partial_{\varphi}\right)  +\sigma_{z}m. \label{Hm}%
\end{equation}
where $\sigma_{x}=\left(
\begin{array}
[c]{lc}%
0 & 1\\
1 & 0
\end{array}
\right)  $, $\sigma_{y}=\left(
\begin{array}
[c]{lc}%
0 & -i\\
i & 0
\end{array}
\right)  $ are, respectively, the $x,y$ -component of Pauli matrices. Now,
whether a geometric potential exists in the relativistic Hamiltonian
(\ref{Hm}) is going to be resolved.

First, let us assume that the most general form of the geometric potential is
given by,%
\begin{equation}
V_{G}=a_{0}I+a_{x}\sigma_{x}+a_{y}\sigma_{y}+a_{z}\sigma_{z} \label{vg}%
\end{equation}
where $\left(  a_{0},a_{x},a_{y},a_{z}\right)  \ $are function of $\theta$ and
$\varphi$. The trial Hamiltonian is now $H^{\prime}=H+V_{G}$.

Secondly, we compute three commutators $\left[  p_{i},H^{\prime}\right]  $ and
the results are, respectively,%
\begin{subequations}
\begin{align}
\left[  p_{x},H^{\prime}\right]   &  =-\hbar^{2}\left(  \sigma_{x}\cos
\varphi\left(  \sin\theta\partial_{\theta}+\cos\theta\right)  +\sigma
_{y}\left(  \cos\varphi\partial_{\varphi}-\frac{\sin\varphi}{2}\right)
\right)  +\left[  p_{x},V_{G}\right]  ,\label{pxh}\\
\left[  p_{y},H^{\prime}\right]   &  =-\hbar^{2}\left(  \sigma_{x}\sin
\varphi\left(  \sin\theta\partial_{\theta}+\cos\theta\right)  +\sigma
_{y}\left(  \sin\varphi\partial_{\varphi}+\frac{\cos\varphi}{2}\right)
\right)  +\left[  p_{y},V_{G}\right]  ,\label{pyh}\\
\left[  p_{z},H^{\prime}\right]   &  =-\hbar^{2}\left(  \sigma_{x}\left(
\cos\theta\partial_{\theta}+\frac{1}{2\sin\theta}-\sin\theta\right)
+\sigma_{y}\frac{\cos\theta}{\sin\theta}\partial_{\varphi}\right)  +\left[
p_{z},V_{G}\right]  , \label{pzh}%
\end{align}
Thirdly, since the commutation relations $\left[  p_{i},V_{G}\right]  $ in
above equation are, respectively,%
\end{subequations}
\begin{subequations}
\begin{align}
\left[  p_{x},V_{G}\right]   &  =-i\hbar\left(  \cos\theta\cos\varphi
\partial_{\theta}V_{G}-\frac{\sin\varphi}{\sin\theta}\partial_{\varphi}%
V_{G}\right)  +i\hbar\frac{\cos\theta\sin\varphi}{\sin\theta}\left(
a_{x}\sigma_{y}-a_{y}\sigma_{x}\right)  ,\label{pxv}\\
\left[  p_{y},V_{G}\right]   &  =-i\hbar\left(  \cos\theta\sin\varphi
\partial_{\theta}V_{G}+\frac{\cos\varphi}{\sin\theta}\partial_{\varphi}%
V_{G}\right)  -i\hbar\frac{\cos\theta\cos\varphi}{\sin\theta}\left(
a_{x}\sigma_{y}-a_{y}\sigma_{x}\right)  ,\label{pyv}\\
\left[  p_{z},V_{G}\right]   &  =i\hbar\sin\theta\partial_{\theta}V_{G},
\label{pzv}%
\end{align}
we immediately have for $x$, $y$, and $z$ -component of the operator
$\mathbf{n}\times\lbrack\mathbf{p},H]-[\mathbf{p},H]\times\mathbf{n}$,
respectively,%
\end{subequations}
\begin{subequations}
\begin{align}
&  2i\hbar\left(  \sin\varphi\partial_{\theta}V_{G}+\frac{\cos\theta
\cos\varphi}{\sin\theta}\partial_{\varphi}V_{G}+\frac{\cos^{2}\theta
\cos\varphi}{\sin\theta}\left(  a_{x}\sigma_{y}-a_{y}\sigma_{x}\right)
\right)  ,\label{xy}\\
&  2i\hbar\left(  -\cos\varphi\partial_{\theta}V_{G}+\frac{\cos\theta
\sin\varphi}{\sin\theta}\partial_{\varphi}V_{G}+\frac{\cos^{2}\theta
\sin\varphi}{\sin\theta}\left(  a_{x}\sigma_{y}-a_{y}\sigma_{x}\right)
\right)  ,\label{yz}\\
&  -2i\hbar\left(  \partial_{\varphi}V_{G}+\cos\theta\left(  a_{x}\sigma
_{y}-a_{y}\sigma_{x}\right)  \right)  . \label{zx}%
\end{align}
In final, DQCs $\mathbf{n}\wedge\lbrack\mathbf{p},H^{\prime}]-[\mathbf{p}%
,H^{\prime}]\wedge\mathbf{n}=0$ would be violated unless $\left(  a_{0}%
,a_{x},a_{y},a_{z}\right)  =(0,0,0,0)$. We see a definite result that there is
not geometric potential, i.e., $V_{G}=0$. In other words, Hamiltonian
(\ref{Hm}) holds true.

\section{Conclusions}

For a particle that is constrained on an ($N-1$)-dimensional curved surface,
the geometric momentum (\ref{ver1}) has stood both theoretical examinations
and experimental testifications. To make it general covariance so as to be
applied to the spin particles, a simple replacement suffices of ordinary
derivative $\partial_{\mu}$ in gradient operator ${\nabla_{\Sigma}\equiv
}\mathbf{e}^{\mu}\partial_{\mu}$ by its general covariant derivative
$\partial_{\mu}\longrightarrow\partial_{\mu}+i\Omega_{\mu}$. A general
formalism when quantizing a classical system is established, and we have FQCs
and DQCs. The FQCs for a Dirac fermion on $S^{N-1}$ lead us to a
\emph{generalized total angular momentum}, while the DQCs for a Dirac fermion
on $\Sigma^{N-1}$ offer us a way to check whether the geometric potential
presents for relativistic spin particle on any hypersurface $\Sigma^{N-1}$. In
present paper, we obtain the \emph{generalized total angular momentum} for
$S^{2}$, which was reported before on the dynamical consideration, and show
that for a Dirac fermion on $S^{2}$, no geometric potential is permissible.
\end{subequations}
\begin{acknowledgments}
This work is financially supported by National Natural Science Foundation of
China under Grant No. 11675051.
\end{acknowledgments}

\end{document}